\documentclass[12pt]{article}
\usepackage{amsmath,amsthm,amsbsy,amsfonts,comment,booktabs}
\usepackage{bm,graphicx,psfrag,epsf,booktabs,bbm,multirow}
\usepackage{enumerate}
\usepackage[round, authoryear]{natbib}
\usepackage{url} % not crucial - just used below for the URL 
\usepackage{ifpdf}
\usepackage[margin=1in]{geometry}
\usepackage{setspace}
\RequirePackage[OT1]{fontenc}
\usepackage{algorithm}
\usepackage{algorithmic}
\usepackage{paralist}

\usepackage{amsbsy}

\usepackage[mathscr]{eucal}

\usepackage{bm}

\usepackage{xspace}

\usepackage{subfig}

\mathchardef\mhyphen="2D

\ifpdf
\usepackage[colorlinks,urlcolor=blue,citecolor=blue,linkcolor=blue]{hyperref}
\else
\newcommand{\href}[2]{{#2}}
\usepackage{url}
\fi

\ifpdf
\newcommand{\Sec}[1]{\hyperref[sec:#1]{Section~\ref*{sec:#1}}} %section
\newcommand{\App}[1]{\hyperref[sec:#1]{Appendix~\ref*{sec:#1}}} %appendix
\newcommand{\Supp}[1]{\hyperref[sec:#1]{supplement~\ref*{sec:#1}}} %supplement
\newcommand{\Eqn}[1]{\hyperref[eq:#1]{{\rm (\ref*{eq:#1})}}} %equation
\newcommand{\Part}[1]{\hyperref[part:#1]{(\ref*{part:#1})}} %part of theorem
\newcommand{\Fig}[1]{\hyperref[fig:#1]{Figure~\ref*{fig:#1}}} %figure
\newcommand{\Tab}[1]{\hyperref[tab:#1]{Table~\ref*{tab:#1}}} %table
\newcommand{\Thm}[1]{\hyperref[thm:#1]{Theorem~\ref*{thm:#1}}} %theorem
\newcommand{\Lem}[1]{\hyperref[lem:#1]{Lemma~\ref*{lem:#1}}} %lemma
\newcommand{\Prop}[1]{\hyperref[prop:#1]{Proposition~\ref*{prop:#1}}} %proposition
\newcommand{\Cor}[1]{\hyperref[cor:#1]{Corollary~\ref*{cor:#1}}} %corollary
\newcommand{\Def}[1]{\hyperref[def:#1]{Definition~\ref*{def:#1}}} %definition
\newcommand{\Alg}[1]{\hyperref[alg:#1]{Algorithm~\ref*{alg:#1}}} %algorithm
\newcommand{\Ex}[1]{\hyperref[ex:#1]{Example~\ref*{ex:#1}}} %example
\newcommand{\As}[1]{\hyperref[as:#1]{Assumption~{\rm\ref*{as:#1}}}} %assumption
\newcommand{\Reg}[1]{\hyperref[as:#1]{Condition~\ref*{reg:#1}}} %regularity condition
\newcommand{\AlgLine}[2]{\hyperref[alg:#1]{line~\ref*{line:#2} of Algorithm~\ref*{alg:#1}}}
\newcommand{\AlgLines}[3]{\hyperref[alg:#1]{lines~\ref*{line:#2}--\ref*{line:#3} of Algorithm~\ref*{alg:#1}}}
\else
\newcommand{\Sec}[1]{{Section~\ref{sec:#1}}} %section
\newcommand{\App}[1]{{Appendix~\ref{sec:#1}}} %appendix
\newcommand{\Supp}[1]{{supplement~\ref{sec:#1}}} %supplement
\newcommand{\Eqn}[1]{{(\ref{eq:#1})}} %equation
\newcommand{\Part}[1]{{(\ref{part:#1})}} %part of proof
\newcommand{\Fig}[1]{{Figure~\ref{fig:#1}}} %figure
\newcommand{\Tab}[1]{{Table~\ref{tab:#1}}} %table
\newcommand{\Thm}[1]{{Theorem~\ref{thm:#1}}} %theorem
\newcommand{\Lem}[1]{{Lemma~\ref{lem:#1}}} %lemma
\newcommand{\Prop}[1]{{Proposition~\ref{prop:#1}}} %proposition
\newcommand{\Cor}[1]{{Corollary~\ref{cor:#1}}} %corollary
\newcommand{\Def}[1]{{Definition~\ref{def:#1}}} %definition
\newcommand{\Alg}[1]{{Algorithm~\ref{alg:#1}}} %algorithm
\newcommand{\Ex}[1]{{Example~\ref{ex:#1}}} %example
\newcommand{\Reg}[1]{{R~\ref*{reg:#1}}} %regularity condition
\fi

\newcommand{\halfquad}{\quad\!\!}
\newcommand{\Real}{\mathbb{R}}

\newcommand{\Tra}{^{\sf T}} % Transpose
\newcommand{\Inv}{^{-1}} % Transpose

\newcommand{\diag}{\mathrm{diag}}

\newcommand{\V}[1]{{\bm{\mathbf{\MakeLowercase{#1}}}}} % vector
\newcommand{\VE}[2]{\MakeLowercase{#1}_{#2}} % vector element
\newcommand{\M}[1]{{\bm{\mathbf{\MakeUppercase{#1}}}}} % matrix
\newcommand{\Mn}[2]{\M{#1}^{(#2)}} % n-th matrix
\newcommand{\prox}[2]{\operatorname{prox}_{#1}({#2})}
\newcommand{\amp}{\mathop{\:\:\,}\nolimits}

\newtheorem{theorem}{Theorem}[section]

\newcommand{\blind}{1}

\begin{document}

\def\spacingset#1{\renewcommand{\baselinestretch}%
{#1}\small\normalsize} \spacingset{1.45}

%%%%%%%%%%%%%%%%%%%%%%%%%%%%%%%%%%%%%%%%%%%%%%%%%%%%%%%%%%%%%%%%%%%%%%%%%%%%%%

\if1\blind
{
  \title{\bf Bayesian Trend Filtering via \\Proximal Markov Chain Monte Carlo}
  \author{\textbf{Qiang Heng\footnote{Department of Statistics, North Carolina State University} \hspace{2mm} Hua Zhou\footnote{Departments of Biostatistics and Computational Medicine, UCLA} \hspace{2mm}  and Eric C.\@ Chi\footnote{Department of Statistics, Rice University}}}
    \date{}
  \maketitle
} \fi

\if0\blind
{
  \bigskip
  \bigskip
  \bigskip
  \begin{center}
    {\LARGE\bf Bayesian Trend Filtering via \\Epigraph-based Proximal MCMC}
\end{center}
  \medskip
} \fi

\begin{abstract}
Proximal Markov Chain Monte Carlo is a novel construct that lies at the intersection of Bayesian computation and convex optimization, which helped popularize the use of nondifferentiable priors in Bayesian statistics. Existing formulations of proximal MCMC, however, require hyperparameters and regularization parameters to be prespecified. In this work, we extend the paradigm of proximal MCMC through introducing a novel new class of nondifferentiable priors called epigraph priors. As a proof of concept, we place trend filtering, which was originally a nonparametric regression problem, in a parametric setting to provide a posterior median fit along with credible intervals as measures of uncertainty. The key idea is to replace the nonsmooth term in the posterior density with its Moreau-Yosida envelope, which enables the application of the gradient-based MCMC sampler Hamiltonian Monte Carlo. The proposed method identifies the appropriate amount of smoothing in a data-driven way, thereby automating regularization parameter selection. Compared with conventional proximal MCMC methods, our method is mostly tuning free, achieving simultaneous calibration of the mean, scale and regularization parameters in a fully Bayesian framework. Supplementary materials for this article are available online.
\end{abstract}

\noindent%
{\it Keywords:} convex optimization, epigraphs, Moreau-Yosida envelope, Hamiltonian Monte Carlo, trend filtering
%\spacingset{1.45} % DON'T change the spacing!

%% ----------------------------------------------------------------------
%% Introduction
%% ----------------------------------------------------------------------
\section{Introduction}
\label{sec:intro}
When analyzing time series data, we are often interested in estimating a slowly varying underlying trend with desired properties such as smoothness and shape restrictions. Smoothness can be achieved by constraining the underlying trend to be piecewise polynomial, while shape restrictions such as  monotonicity and convexity can be enforced by linear inequality constraints. Let $\V{y}\in\Real^n$ denote an observed time series  and $\V{\beta}\in\Real^n$ denote its underlying trend; then estimating $\V{\beta}$ is commonly posed as the following constrained or penalized least squares problem
\begin{eqnarray}\label{eq:nonparameteric-problem}
    \underset{\V{\beta}\in \Real^n}{\text{minimize}}& \frac{1}{2}\lVert \V{y}-\V{\beta}\rVert_2^2+g(\V{\beta}),
\end{eqnarray}
where $g(\V{\beta})$ is an indicator function encoding convex constraints or a nonsmooth penalty function inducing sparsity. Different choices of $g(\V{\beta})$ induce a variety of sequence approximation problems. Representative examples include isotonic regression \citep{barlow1972statistical}, univariate convex regression \citep{groeneboom2008support}, nearly-isotonic regression \citep{tibshirani2011nearly} and $\ell_1$-trend filtering \citep{steidl2006splines,kim2009ell_1,tibshirani2014adaptive}. 

As a nonparametric regression problem, the solution to \Eqn{nonparameteric-problem} only produces a point estimate. If we are interested in uncertainty quantification, data-resampling techniques like the bootstrap \citep{efron1994introduction} can be adopted. The bootstrap, however, does not address the issue of regularization parameter selection. The bootstrap is only able to produce a confidence band with a given regularization parameter, which is often selected with cross validation.

To quantify uncertainty and automate regularization parameter selection, many have placed \Eqn{nonparameteric-problem} in a Bayesian framework. Inspired by the Bayesian Lasso \citep{park2008bayesian}, \citet{roualdes2015bayesian} first introduced Bayesian Trend Filtering (BTF), exploiting the Gaussian mixture representation of the Laplace prior. Independent from Rouadle's work, \cite{faulkner2018locally} proposed a closely related smoothing method, Shrinkage Prior Markov Random Fields (SPMRFs), which places sparsity inducing shrinkage priors on the adjacent differences of the elements of $\V{\beta}$. In addition to the Laplace prior, \cite{faulkner2018locally} also investigated a more aggressive horseshoe prior \citep{carvalho2010horseshoe}, which demonstrated superior local adaptivity to abrupt changes or jumps. Recently, \cite{kowal2019dynamic} proposed dynamic shrinkage processes (DSP) for Bayesian trend filtering with even stronger localized adaptivity to irregular features through modelling dependence between the local scale parameters. 

The literature of Bayesian shape-restricted regression is vast and diverse. Early works include Bayesian isotonic regression with piecewise linear models \citep{neelon2004bayesian}, Bayesian P-splines \citep{brezger2008monotonic}, Bayesian monotone regression with Bernstein polynomials \citep{mckay2011variable}. Two more recent methods are Bayesian shape-restricted splines \citep{meyer2011bayesian} and Bayesian shape-restricted regression using Gaussian process priors \citep{lenk2017bayesian}, which can enforce both monotonicity and convexity. 

Our approach to Bayesian trend filtering takes advantage of a relatively new Markov chain Monte Carlo (MCMC) sampling scheme in the Bayesian imaging literature, namely the proximal MCMC methods \citep{pereyra2016proximal,durmus2018efficient,pereyra2020accelerating}. The current paradigm of proximal MCMC methods requires  variance and regularization parameters to be fixed and predetermined. In this work, we incorporate those parameters into posterior inference, leveraging the data itself to automatically determine the appropriate amount of smoothing. We present two applications of our proposed methodology, namely Proximal Bayesian Trend Filtering (PBTF) and Proximal Bayesian Shape-Restricted Trend Filtering (PBSRTF).

%% ----------------------------------------------------------------------
%% Methods
%% ----------------------------------------------------------------------
\section{Background}

We first review the nonparameteric function estimation with $\ell_1$-trend filtering as well as important concepts from convex optimization needed to develop our Bayesian trend filtering algorithms.

\subsection{Nonparametric Estimation with $\ell_1$-trend filtering}\label{sec:ntf}
Suppose that a time series $\V{y}\in \mathbb{R}^n$ observed over a grid of time points $\V{x}\in \mathbb{R}^n$ is the superposition of a smooth trend $\V{\beta}\in \mathbb{R}^n$ and Gaussian noise $\V{\epsilon}\sim\mathcal{N}(0,\sigma^2\M{I}_n)$, namely
\begin{eqnarray}\label{eq:ntfassumption}
\VE{y}{i} & = & \VE{\beta}{i} +\epsilon_i,\quad i = 1,2,\dots,n,
\end{eqnarray}
where the grid locations $x_i$ are strictly increasing, i.e., $x_1<x_2<\dots<x_n$. For simplicity, we assume for now that a single measurement is observed at each grid point and the grid points are evenly spaced. We relax both assumptions later.

\cite{kim2009ell_1} proposed $\ell_1$-trend filtering to estimate $\V{\beta}$ with piecewise polynomial structure, by solving the following regularized least squares problem
\begin{eqnarray}\label{eq:ntfobjective}
    \underset{\V{\beta}\in \Real^n}{\text{minimize}}& \frac{1}{2}\lVert \V{y}-\V{\beta}\rVert_2^2+\alpha\lVert \M{D}^{(k+1)}_n\V{\beta} \rVert_1,
\end{eqnarray}
where $\alpha$ is a positive regularization parameter,  $\M{D}^{(k+1)}_n\in \mathbb{R}^{(n-k-1)\times n}$ is the discrete difference operator or matrix of order $k+1$ and dimension $n$. To appreciate the effect of penalizing the $\ell_1$-norm of $ \M{D}^{(k+1)}_n\V{\beta}$, we explicitly write out the difference operator for $k=0$,
\begin{eqnarray*}
    \M{D}^{(1)}_n & = &
\begin{bmatrix}
-1 & 1 & 0 & \dots & 0 & 0\\
0 & -1 & 1 & \dots & 0 & 0\\
\vdots & \vdots & \ddots & \ddots & \vdots & \vdots \\
0 & 0 & \dots &-1 & 1 & 0 \\
0 & 0 & \dots & 0 & -1 &1
\end{bmatrix}\in\Real^{(n-1)\times n}.
\end{eqnarray*}
When $k=0$, the penalty term $\lVert \M{D}^{(1)}_n\V{\beta}\rVert_1=\sum_{i=1}^{n-1}\lvert \beta_{i+1}-\beta_i\rvert$ is also known as the one-dimensional total variation denoising penalty \citep{rudin1992non,steidl2006splines} in signal processing, or the fused lasso penalty \citep{tibshirani2005sparsity} in statistics. The penalty incentivizes recovery of  piecewise constant solutions. Higher-order difference matrices are defined recursively as $\M{D}^{(k+1)}_n = \M{D}^{(1)}_{n-k}\M{D}^{(k)}_n.$ 
Choosing order $k=1,2,$ and $3$ incentivizes the recovery of piecewise linear, quadratic and cubic solutions, respectively. Difference matrices of order higher than 4 are rarely of interest.

To handle irregular grids, namely when the time points $\V{x}\in\Real^n$ are strictly increasing but possibly unevenly spaced, \cite{tibshirani2014adaptive} proposed replacing $\M{D}^{(k+1)}_n$ with the adjusted difference matrix $\M{D}^{(\V{x},k+1)}_n$. The first-order difference matrix remains the same, i.e. $\M{D}^{(\V{x},1)}_n=\M{D}^{(1)}_n$; for $k\ge1$ the adjusted difference operators are now defined as
\begin{eqnarray*}
\M{D}^{(\V{x},k+1)}_n &=& \M{D}^{(\V{x},1)}_{n-k}\operatorname{diag}\left(\frac{k}{x_{k+1}-x_1},\dots,\frac{k}{x_{n}-x_{n-k}}\right) \M{D}^{(\V{x},k)}_n\; \text{ for }\;k = 1, 2, \dots
\end{eqnarray*}
Note when $x_1=1,x_2=2,\dots,x_n=n$, the adjusted difference matrix $\M{D}^{(\V{x},k+1)}_n$ coincides with $\M{D}^{(k+1)}_n$. 

A variety of iterative and non-iterative algorithms have been proposed to compute a solution to \Eqn{ntfobjective}. The ones that are relevant to this work are the dynamic programming algorithm by \cite{johnson2013dynamic} and the ADMM algorithm by \cite{ramdas2016fast}. Remarkably, the dynamic programming approach can solve \Eqn{ntfobjective} exactly in $O(n)$ steps for $k=0$. Building on top of the dynamic programming algorithm, the ADMM algorithm solves \Eqn{ntfobjective} iteratively for $k=1,2,$ and 3. 

As discussed in \cite{kim2009ell_1}, adding additional shape restrictions to $\ell_1$-trend filtering is straightforward. For example, one might require the underlying trend to be monotone-increasing. The isotonic $\ell_1$-trend filtering problem is formulated as
\begin{eqnarray*}
\underset{\V{\beta}\in \Real^n}{\text{minimize}}& \frac{1}{2}\lVert \V{y}-\V{\beta}\rVert_2^2+\alpha\lVert \M{D}^{(\V{x},k+1)}_n\V{\beta} \rVert_1 \quad \text{subject to} \quad \beta_1\le\beta_2\le \dots \le \beta_n.
%\text{s.t. } & \beta_1\le\beta_2\le \dots \le \beta_n.
\end{eqnarray*}
The monotonicity constraint $\beta_1\le\beta_2\le \dots \le \beta_n$ can be written compactly as $\M{D}^{(1)}_n\V{\beta}\ge \V{0}$, where $\ge$ represents elementwise inequality.

In addition to monotonicity, another common shape restriction is convexity. The underlying trend $\V{\beta}$ is convex if
\begin{eqnarray}\label{eq:convexity}
\frac{\beta_2-\beta_1}{x_2-x_1} & \le & \frac{\beta_3-\beta_2}{x_3-x_2} \amp \le \amp \dots \amp\le\amp \frac{\beta_n-\beta_{n-1}}{x_n-x_{n-1}},
\end{eqnarray}
which can be written compactly as $\M{D}^{(\V{x},2)}_n\V{\beta}\ge \V{0}$.  

For the rest of this paper, we will work with the general case where we may have mulitple observations per grid point. We assume that observations $y_{ij}$ come from the model
\begin{eqnarray}%\label{eq:assumption}
   y_{ij} & = & \beta(x_i)+\epsilon_{ij},\;\epsilon_{ij}\overset{i.i.d.}\sim\mathcal{N}(0,\sigma^2),\;i=1,2,\dots,n,\;j=1,2,\dots,w_i,
\end{eqnarray}
where $\beta(x)$ is the underlying trend function that we seek to estimate and $w_i$ is the number of observations at a particular grid location $x_i$. We assume that the underlying function $\beta(x)$ has piecewise polynomial structure. Allowing multiple observations at a given grid location is useful as real data is often discrete.

\subsection{Relevant Concepts from Convex Optimization}
We next review concepts from convex optimization central to our proposed framework, specifically projection and proximal mappings which are the algorithmic primitives that we will use to build our Bayesian trend filtering methods.

In convex analysis, the indicator function $\iota_\mathcal{A}(\V{\beta})$ of a set $\mathcal{A} \subset \mathbb{R}^n$ takes on the value of 0 when $\V{\beta} \in \mathcal{A}$ and the value of $+\infty$ when $\V{\beta} \notin \mathcal{A}$. The familiar 0-1 indicator function $\mathbbm{1}_\mathcal{A}(\V{\beta})$, which takes on the value of 1 when $\V{\beta} \in \mathcal{A}$ and 0 when $\V{\beta} \notin \mathcal{A}$ is an invertible transformation the indicator function from convex analysis, namely $\mathbbm{1}_\mathcal{A}(\V{\beta}) = \exp(-\iota_\mathcal{A}(\V{\beta}))$.
The projection of a point $\V{\beta}$ onto a set $\mathcal{A}$, denoted by $P_{\mathcal{A}}(\V{\beta})$, is a point in $\mathcal{A}$ that is closest in Euclidean distance to $\V{\beta}$.
\begin{eqnarray*}
P_{\mathcal{A}}(\V{\beta}) & = & \underset{\V{\eta}\in \mathcal{A}}{\arg\min}\; \lVert \V{\eta}-\V{\beta}\rVert_2.
\end{eqnarray*}
The projection $P_{\mathcal{A}}(\V{\beta})$ exists and is unique when $\mathcal{A}$ is closed and convex,.

The proximal map of the function $g$ is the following operator
\begin{eqnarray*}
 \operatorname{prox}_g(\V{\beta}) & = & \underset{\V{\eta} \in \Real^n}{\operatorname{argmin}}\;\left[g(\V{\eta})+\frac{1}{2} \lVert \V{\beta} - \V{\eta} \rVert_2^2 \right].
\end{eqnarray*}
An additional positive parameter $\lambda$ is often added to control proximity,
\begin{eqnarray*}
 \operatorname{prox}_{\lambda g}(\V{\beta}) & = & \underset{\V{\eta} \in \Real^n}{\operatorname{argmin}}\;\left[g(\V{\eta})+\frac{1}{2\lambda} \lVert \V{\beta} - \V{\eta} \rVert_2^2 \right].
\end{eqnarray*}
Following the notation in prior proximal MCMC papers, we write $\operatorname{prox}_{\lambda g}(\V{\beta})$ as $\operatorname{prox}_{g}^\lambda(\V{\beta})$. 

When $g$ is an indicator function of a set $\mathcal{A}$, the proximal operator is the projection onto $\mathcal{A}$. Consequently, proximal maps generalize projection operations. Proximal maps play an important role in modern machine learning due to the fact that many nonsmooth penalties often have unique proximal mappings that either have explicit formulas or can be computed efficiently. In this work, we take advantage of two such proximal mappings, namely the proximal maps of $\lVert \V{\beta}\rVert_1$ and $\lVert \M{D}^{(1)}_n\V{\beta}\rVert_1$. The proximal map of $\lVert \V{\beta} \rVert_1$ is the celebrated soft-threshold operator
\begin{eqnarray}\label{eq:softthreshold}
\left [\operatorname{prox}^{\lambda}_{g}(\V{\beta})\right]_i &= &\begin{cases}
\VE{\beta}{i} & \lvert\VE{\beta}{i}\vert \le \lambda\\
\operatorname{sgn}(\VE{\beta}{i}) (\lvert \VE{\beta}{i} \rvert - \lambda)_+ &  \VE{\beta}{i}\rvert > \lambda
\end{cases},
\end{eqnarray}
while the proximal map of $\lVert \M{D}^{(1)}_n\V{\beta} \rVert_1$ is the solution to the fused Lasso problem \citep{tibshirani2005sparsity}:
\begin{eqnarray}\label{eq:fusedlasso}
\operatorname{prox}^{\lambda}_g(\V{\beta}) &=&\underset{\V{\eta}\in\mathbb{R}^n}{\operatorname{argmin}}\;\frac{1}{2}\lVert \V{\beta}-\V{\eta}\rVert_2^2+\lambda \lVert \M{D}^{(1)}_n\V{\eta} \rVert_1,
\end{eqnarray}
which can be solved exactly in linear time via dynamic programming  \citep{johnson2013dynamic}.
We use these two proximal maps as a subroutine to perform a key computation, namely the epigraph projection, which we will describe later. 

The $\lambda$-Moreau-Yosida envelope of a function $g$ is given by 
\begin{eqnarray*}
g^\lambda(\V{\beta}) &= &\underset{\V{\eta}\in \Real^n}\min\; g(\V{\eta})+\frac{1}{2\lambda}\lVert \V{\eta} - \V{\beta}\rVert_2^2.
\end{eqnarray*}
The envelope function $g^\lambda$ has several important properties. First, $g^\lambda$ is convex when $g$ is convex. Second, $g^\lambda$ is always differentiable even if $g$ is not, and its gradient can be expressed in terms of the proximal map of $\lambda g$, namely
\begin{eqnarray*}
\nabla g^\lambda (\V{\beta}) & = & \frac{1}{\lambda}\left[\V{\beta}-\operatorname{prox}^\lambda_g(\V{\beta}) \right].
\end{eqnarray*}
Moreover, $\nabla g^\lambda$ is $\lambda\Inv$-Lipschitz since proximal operators are firmly nonexpansive \citep{combettes2011proximal}. Finally and perhaps most importantly,  $g^\lambda$ converges pointwise to $g$ as $\lambda$ tends to 0~\citep{rockafellar2009variational}. In short, we see that the Moreau-Yosida envelope of a nonsmooth function $g$ is a  Lipschitz-differentiable, arbitrarily close approximation to $g$. In this work, we will rely on the Moreau-Yosida envelope of indicator functions. Since the proximal map of an indicator function $\iota_{\mathcal{E}}(\V{\beta})$ is the projection $P_{\mathcal{E}}(\V{\beta})$, its Moreau-Yosida envelope is
$ g^\lambda (\V{\beta}) = \frac{1}{2\lambda}\lVert \V{\beta}- P_{\mathcal{E}}(\V{\beta})\rVert_2^2$,
where $\lVert \V{\beta}- P_{\mathcal{E}}(\V{\beta})\rVert_2$ is also denoted as $d_{\mathcal{E}}(\V{\beta})$, namely the distance of $\V{\beta}$ to $\mathcal{E}$. 

The Moreau-Yosida approximation is the key technical ingredient behind the proximal MCMC framework of~\cite{durmus2018efficient} which our algorithmic framework extends. We next review their prior formulation of the proximal MCMC method.

\section{Proximal MCMC}\label{sec:proximalmcmc}

Many modern machine learning applications employ log-concave models of the form
\begin{eqnarray}
\label{eq:original}
\pi(\V{\beta}) & \propto & \exp\{-U(\V{\beta})\} \quad\quad\text{and}\quad\quad U(\V{\beta}) \amp = \amp f(\V{\beta}) + g(\V{\beta}),
\end{eqnarray}
where $f$ is a Lipschitz-differentiable convex negative log-likelihood function and $g$ is a lower-semicontinuous convex penalty function that shrinks the estimator towards some desired prior structure. The model in \Eqn{ntfassumption} that underlies the $\ell_1$-trend-filtering problem is an example of such a log-concave model, where
\begin{eqnarray*}
f(\V{\beta}) & = & \frac{1}{2\sigma^2}\lVert \V{y} - \V{\beta} \rVert_2^2 \quad\quad\text{and}\quad\quad
g(\V{\beta}) \amp = \amp \alpha \lVert \Mn{D}{k+1}_n\V{\beta} \rVert_1.
\end{eqnarray*}
Note that if we absorb $\sigma^2$ into the regularization parameter $\alpha$, then computing the maximum a posteriori (MAP) estimate of $\V{\beta}$ in this log-concave model is equivalent to solving the nonparameteric problem \Eqn{ntfobjective}.

Given such a log-concave model, we may wish to facilitate uncertainty quantification and posterior inference by computing posterior samples. Unfortunately, while there are many scalable methods for computing the MAP estimate of $\V{\beta}$, for example the Split-Bregman~\citep{GoldsteinOsher2009} and Chambolle-Pock~\citep{ChambollePock2011} algorithms, 
sampling from the posterior distribution \Eqn{original} is not as straightforward. Conventional high-dimensional MCMC algorithms, such as the unadjusted Langevin algorithm (ULA)~\citep{roberts1996exponential}, Metropolis-adjusted Langevin algorithm (MALA)~\citep{rossky1978brownian,roberts1996exponential}, Hamiltonian Monte Carlo (HMC)~\citep{neal2011mcmc}, rely on gradient mappings that in turn require $U$ to be Lipschitz-differentiable or at least differentiable.  These differentiability requirements can be extremely limiting, as they rule out many commonly used nonsmooth penalty functions $g$.

To make efficient high-dimensional MCMC algorithms applicable for nonsmooth $U$, \cite{pereyra2016proximal} proposed replacing $U$ with a Lipschitz-differentiable approximation, namely the $\lambda$-Moreau-Yosida envelope of $U$, and then employing MALA to sample from the derived surrogate density (Px-MALA).  \cite{durmus2018efficient} proposed a slightly different strategy with the Moreau-Yosida regularized Unadjusted Langevin Algorithm (MYULA), by replacing $g$ with its Moreau-Yosida approximation $g^\lambda$ in \Eqn{original} to obtain the surrogate density 
\begin{eqnarray}
\label{eq:surrogate}
\pi^\lambda(\V{\beta}) & \propto & \exp\{- f(\V{\beta}) - g^\lambda(\V{\beta})\}.
\end{eqnarray}
Under additional assumptions on $g$, the surrogate density \Eqn{surrogate} is proper and converges to the original density \Eqn{original} in total-variation norm~\citep{durmus2018efficient}. Moreover, if $g$ is Lipschitz, then the total-variation norm of \Eqn{original} and \Eqn{surrogate} is bounded linearly in $\lambda$. The MYULA algorithm simply applies ULA to the surrogate density \Eqn{surrogate}:
\begin{eqnarray}\label{eq:MYULAtransition}
\V{\beta}_{l+1} &= &\left(1-\frac{\gamma}{\lambda}\right)\V{\beta}_{l}-\gamma\nabla f(\V{\beta}_{l})+\frac{\gamma}{\lambda}\operatorname{prox}_g^\lambda(\V{\beta}_l)+\sqrt{2\gamma}\V{\zeta}_{l+1},
\end{eqnarray}
where $\V{\zeta}_{l+1}$ is $n$-dimensional Brownian motion and $\gamma$ is the step size of ULA. A Metropolis-Hastings correction step can be added to remove the asymptotic bias associated with Euler-Maruyama discretization that is common to Langevin algorithms. An extension of the MYULA algorithm is to combine several gradient evaluations to accelerate its convergence (SK-ROCK) \citep{pereyra2020accelerating}. The recent review paper \cite{durmus2022proximal} provides an overview for proximal MCMC methods and their applications in imaging inverse problems.

\begin{figure}[htbp]
  \centering
  \includegraphics[width=\linewidth]{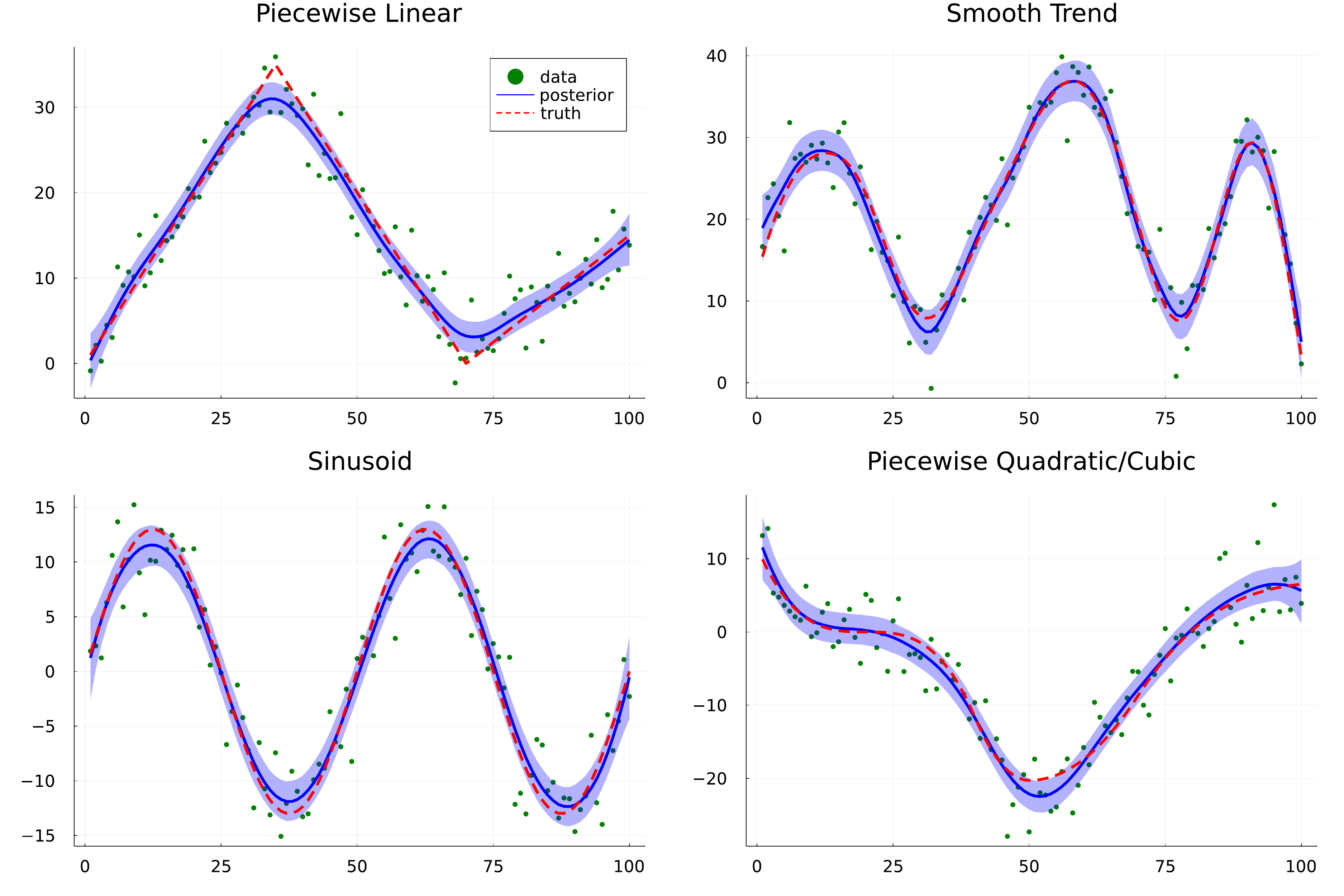}
  \caption{Example posterior fits for PBTF with noise level $\sigma=3$. The standard deviation of the underlying trends is $9$, thus the signal-to-noise ratio is 3. Plots show data points (green dots), posterior median (blue solid lines), 95\% Bayesian credible intervals (light blue bands) and true trends (red dashed lines). }
  \label{fig:pbtfimage}
\end{figure}

A hallmark application of proximal MCMC is Bayesian image deblurring,  where $\V{\beta}$ is a high-dimensional latent image, $f$ is the negative log-likelihood that models blurring and additive Gaussian noise that together corrupt the latent image, and $g$ is a total variation penalty that incentivizes the recovery of a latent image with sharp edges~\citep{durmus2018efficient,pereyra2020accelerating,durmus2022proximal}. In this context, the posterior of interest is
\begin{eqnarray}
\label{eq:tvnormdenoise}
   \pi(\V{\beta} \mid \V{y}) & \propto & \exp\left\{-\frac{\lVert \V{y}-\M{H}\V{\beta} \rVert_2^2}{2\sigma^2}-\alpha \operatorname{TV}(\V{\beta})\right\},
\end{eqnarray}
where $\M{H}$ is a blur operator, $\operatorname{TV}(\V{\beta})$ is the total-variation semi-norm of $\V{\beta}$~\citep{chambolle2004algorithm}, $\V{y}$ is the corrupted image signal we observe, $\sigma^2$ is the noise variance, and $\alpha$ is a positive regularization parameter that trades off the emphasis between data fit and smoothness in the estimated image.
In the framework of \cite{durmus2018efficient} and \citet{pereyra2020accelerating}, the variance $\sigma^2$ and the regularization parameter $\alpha$ need to be manually selected by an expert or determined by an empirical Bayesian method \citep{vidal2020maximum,de2020maximum}. In this work, we propose to use a new construct that we refer to as epigraph priors and HMC sampling to incorporate $\sigma^2$ and $\alpha$ into posterior inference in the context of Bayesian trend filtering. Consequently, this work demonstrates how  proximal MCMC can be applied as a statistical methodology in a unified and complete Bayesian framework. \Fig{pbtfimage} illustrates four examples of posterior fits using our fully Bayesian proximal MCMC method for trend filtering.

%% ----------------------------------------------------------------------
%% Methods
%% ----------------------------------------------------------------------
%\section{Method}\label{sec:methods}

%% ----------------------------------------------------------------------
%% Proximal Bayesian Trend Filtering
%% ----------------------------------------------------------------------
\newpage
 \section{Proximal Bayesian Trend Filtering}\label{sec:methods}
Our key methodological innovation that enables extending the proximal MCMC framework to a fully Bayesian one is the use of epigraph indicator functions to encode our structure-inducing prior. Prior proximal MCMC methods typically replace a nonsmooth penalty $g(\V{\beta})=\alpha h(\V{\beta})$ with its Moreau envelope in the posterior. The proximal operator is then evaluated as 
\begin{eqnarray*}
\operatorname{prox}^\lambda_g(\V{\beta}) & = & \operatorname{prox}^{\lambda \alpha}_h(\V{\beta}),
\end{eqnarray*}
where the proximal operator of $h$ can be computed with an efficient off-the-shelf algorithm. The gradient of $g^\lambda(\V{\beta})$ can then be computed as $(\V{\beta}-\operatorname{prox}^\lambda_g(\V{\beta}))/\lambda$, which is a well-known fact about Moreau envelopes. However, the regularization parameter $\alpha$ is viewed as a hyperparameter in $g^\lambda$ and needs to be determined prior to MCMC sampling. Although an empirical Bayesian method \citep{vidal2020maximum,de2020maximum} can be used to estimate the appropriate $\alpha$ and $\sigma^2$, a fully Bayesian treatment is desirable since it may have better precision due to being able to account for the uncertainty of  $\alpha$ and $\sigma^2$.
\begin{figure}[htbp]
  \centering
  \includegraphics[width=0.8\linewidth]{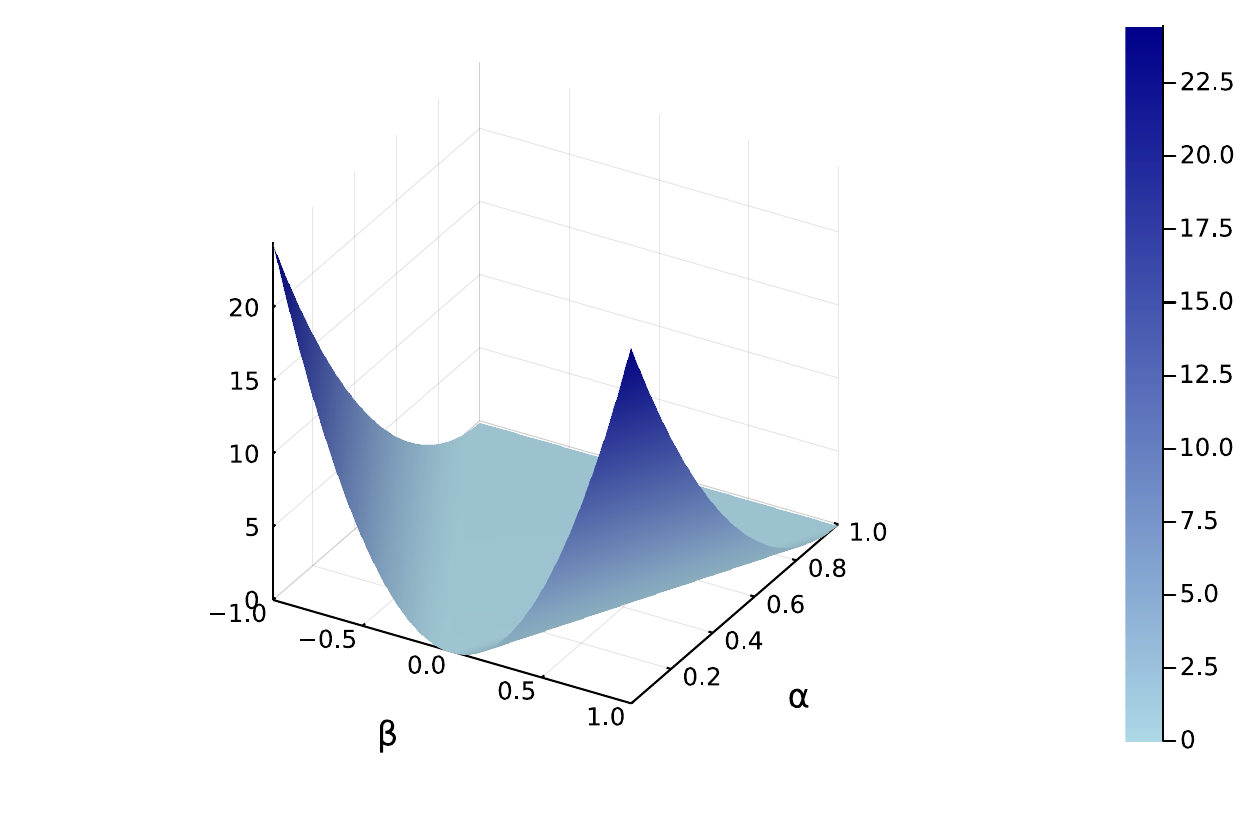}
  \caption{A visualization of the distance function $\frac{1}{2\lambda}d^2_\mathcal{E}(\V{\beta},\alpha)$ when $\mathcal{E}=\{(\beta,\alpha)\in \mathbb{R}^2 : |\beta|\le\alpha\}$ and $\lambda=0.01$.}
  \label{fig:epi_dist}
\end{figure}

To incorporate $\alpha$ into posterior inference, an important concept in convex analysis, epigraph, comes in handy. The epigraph of a regularization function $g$ is the set
\begin{eqnarray*}
\mathcal{E} & = & \{(\V{\beta},\alpha)\in \Real^n \times \Real : g(\V{\beta})\le\alpha\}.
\end{eqnarray*}
The Moreau-Yosida envelope of $\iota_\mathcal{E}(\V{\beta},\alpha)$ is $\frac{1}{2\lambda}d^2_\mathcal{E}(\V{\beta},\alpha)$, which is jointly differentiable in $\V{\beta}$ and $\alpha$. The gradient of $\frac{1}{2\lambda}d^2_\mathcal{E}(\V{\beta},\alpha)$ is simply $\frac{(\V{\beta},\alpha)-P_{\mathcal{E}}(\V{\beta},\alpha)}{\lambda}$, where $P_{\mathcal{E}}$ denotes projection on to $\mathcal{E}$. \Fig{epi_dist} provides a visualization of the envelope function  $\frac{1}{2\lambda}d^2_\mathcal{E}(\V{\beta},\alpha)$ when $\mathcal{E}=\{(\beta,\alpha)\in \mathbb{R}^2 : |\beta|\le\alpha\}$ and $\lambda=0.01$. Using $\frac{1}{2\lambda}d^2_\mathcal{E}(\V{\beta},\alpha)$ as our prior regularization term, we can further place hyperpriors on $\alpha$, $\sigma^2$ and achieve fully Bayesian inference within the proximal MCMC framework. Computing with these priors relies on projection onto epigraphs which we describe next. 

\subsection{Projection Onto Epigraph}\label{sec:poe}

Projection onto the epigraph of $g$ depends on the proximal mapping of $g$ (see Theorem 6.36 of \cite{beck2017first}), namely
\begin{eqnarray}\label{eq:epiprojection}
P_{\operatorname{epi}(g)}(\V{\beta},\alpha) &=&\begin{cases}
(\V{\beta},\alpha) & g(\V{\beta})\le\alpha\\
\left(\operatorname{prox}^{\lambda^*}_g(\V{\beta}),\alpha+\lambda^* \right) & g(\V{\beta})>\alpha
\end{cases},
\end{eqnarray}
where 
 $\lambda^*$ is root of the auxiliary function
\begin{eqnarray*}
F(\lambda) & = & g\left(\operatorname{prox}^{\lambda}_g(\V{\beta})\right)-\lambda-\alpha.
\end{eqnarray*}
When $\operatorname{prox}^{\lambda}_g(\V{\beta})$ can be computed easily, we can compute the root $\lambda^*$
of the function $F(\lambda)$ using a simple bisection procedure. 

We will need to perform projections onto two sets: the epigraph of the $\ell_1$-norm
\begin{eqnarray*}
\mathcal{E}_1 &=&\{(\V{\beta},\alpha)\in\Real^n\times\Real_{++} : \lVert \V{\beta} \rVert_1\le\alpha\},
\end{eqnarray*}
and the epigraph of $\lVert \M{D}^{(1)}_n\V{\beta} \rVert_1$
\begin{eqnarray*}
\mathcal{E}_2 &=& \left\{(\V{\beta},\alpha)\in\Real^n\times\Real_{++} : \lVert \M{D}^{(1)}_n\V{\beta} \rVert_1\le\alpha \right\}.
\end{eqnarray*}
Since the proximal maps of $\lVert \V{\beta} \rVert_1$ and $\lVert \Mn{D}{1}_n\V{\beta}\rVert_1$ can be computed in linear time, projections onto $\mathcal{E}_1$ and $\mathcal{E}_2$ can be done efficiently. For projection onto $\mathcal{E}_1$, we set the initial bisection interval to be $(0,\lambda_{\max})$ where $\lambda_{\max} = \lVert \V{\beta} \rVert_\infty$ is the smallest value of $\lambda$ such that $\prox{\lambda\lVert \cdot \rVert_1}{\V{\beta}} = \V{0}$. For projection onto $\mathcal{E}_2$, we set the initial bisection interval to be $(0,\lambda_{\max})$ where
\begin{eqnarray*}
\lambda_{\max} & = & \left\lVert \left [\M{D}^{(1)}_n(\M{D}^{(1)}_n)\Tra\right]\Inv\M{D}^{(1)}_n\V{\beta} \right\rVert_{\infty},
\end{eqnarray*}
is the smallest value of $\lambda$ such that the solution to \Eqn{fusedlasso} is a multiple of the all ones vector. It is easy to  verify that $F(0)>0$ when $(\V{\beta},\alpha)\notin \operatorname{epi}(g)$ and $F(\lambda_{\max})<0$ so that the root of the auxiliary function is guaranteed to lie within $(0,\lambda_{\max})$. 

In a manner akin to \cite{ramdas2016fast}, projecting onto $\mathcal{E}_2$ instead of projecting onto $\mathcal{E}_1$ alleviates numerical issues associated with solving an ill-conditioned linear system, since it enables us to work with a transformation matrix that is one ``order" lower. We will elaborate on this claim in \Sec{pbtf}.

\subsection{Priors for Proximal Bayesian Trend Filtering}\label{sec:pbtf}

To obtain posterior trends with approximate piecewise polynomial structure, we place a constrained ``flat" prior on $\V{\beta}$ to induce sparsity and regularity, namely
\begin{eqnarray}\label{eq:conditionalprior}
    \pi(\V{\beta} \mid \alpha) &= &\alpha^{-(n-k-1)}\exp\left \{-\iota_{\mathcal{E}}(\V{\beta},\alpha)\right\},
\end{eqnarray}
where
\begin{eqnarray*}
\mathcal{E} &= & \left\{(\V{\beta},\alpha)\in\Real^n\times \Real_{++} : \lVert \M{D}^{(\V{x},k+1)}_n\V{\beta}\rVert_1\le \alpha \right\}.
\end{eqnarray*}
Note that implicitly $\alpha$ must be positive in \Eqn{conditionalprior} and all our subsequent equations. The term $\alpha^{-(n-k-1)}$ reflects the fact that
we are constraining $\M{D}^{(\V{x},k+1)}_n\V{\beta}$ to an $(n-k-1)$-dimensional $\ell_1$-norm ball, which has volume proportional to $\alpha^{n-k-1}$. To complete the model specification, we need to place additional priors on $\sigma^2$ and $\alpha$. For $\sigma^2$, the standard inverse Gamma prior $\operatorname{IG}(s,r)$ suffices as the parameters $s$ and $r$ minimally influence the posterior for small values. In contrast, some care is warranted for choosing the prior for $\alpha$. Ideally, we seek a prior that cancels the term $\alpha^{-(n-k-1)}$ to ensure a proper surrogate posterior density. 

A natural strategy is to use a Gamma prior, which achieves the goal of cancelling out $\alpha^{-(n-k-1)}$. Placing a $\Gamma(n-k,\mu)$ prior on $\alpha$, the joint prior on $(\V{\beta},\alpha)$ becomes 
\begin{eqnarray}\label{eq:jointgamma}
    \pi(\V{\beta},\alpha) &=&\exp\{-\iota_{\mathcal{E}}(\V{\beta},\V{\alpha})-\mu\alpha\}.
\end{eqnarray}
Choosing a Gamma prior, however, requires us to choose large $\mu$ values to impose a meaningful amount of shrinkage, which makes $\Gamma(n-k,\mu)$ an informative prior since its variance is $(n-k)/\mu^2$. In that case selecting an appropriate $\mu$ becomes challenging and stymies our goal of operating within a fully Bayesian framework.

Given these challenges with a Gamma prior, we propose using a beta-prime prior. A beta-prime distribution, denoted as $\beta'(s_1,s_2)$, has density
\begin{eqnarray*}
\pi(\alpha) &\propto& \alpha^{s_1-1}(1+\alpha)^{-s_1-s_2}.
\end{eqnarray*}
If we place a $\beta'(n-k,s_2)$ prior on $\alpha$, the joint prior for $(\V{\beta},\alpha)$ becomes 
\begin{eqnarray}\label{eq:jointbetaprime}
\pi(\V{\beta},\alpha)&\propto&\exp\{-\iota_{\mathcal{E}}(\V{\beta},\V{\alpha})-(n-k+s_2)\log(1+\alpha)\}.
\end{eqnarray}
A $\beta'(s_1,s_2)$ distribution has mean $\frac{s_1}{s_2-1}$ and variance $\frac{s_1(s_1+s_2-1)}{(s_2-2)(s_2-1)^2}$. Consequently when $s_2$ is relatively small, the prior has high variance and becomes uninformative. What makes this prior setup preferred over the one induced by the Gamma prior in \Eqn{jointgamma} is that even when  $s_2$ is small, we still have $-(n-k+s_2)\log(1+\alpha)$ as a strong penalty to impose a useful measure of shrinkage. Therefore the beta-prime prior is better than the Gamma prior in terms of hyperparameter sensitivity. Nonetheless, we will revisit using the Gamma prior later as it is better suited for our second application PBSRTF. Why that is the case will be discussed in \Sec{pbsrtf}.

Placing an $\operatorname{IG}(s,r)$ prior on $\sigma^2$ and a $\beta'(n-k,s_2)$ prior on $\alpha$, our full posterior density reads
\begin{align}\label{eq:fullposteriorpbtf}
\begin{split}
    \pi(\V{\beta},\sigma^2,\alpha \mid y)\halfquad\propto\halfquad&(\sigma^2)^{-\frac{m}{2}-s-1}\exp\bigg\{-\frac{\sum_{i=1}^n\sum_{j=1}^{w_i} (y_{ij}-\beta_i)^2+2r}{2\sigma^2}\\ &-\iota_{\mathcal{E}}(\V{\beta},\V{\alpha})-(n-k+s_2)\log(1+\alpha)\bigg\},
\end{split}
\end{align}
where $m=\sum_{i=1}^n w_i$ is the total number of observations. We can rewrite \Eqn{fullposteriorpbtf} in a vectorized format 
\begin{align}\label{eq:fullposteriorvec}
\begin{split}
    \pi(\V{\beta},\sigma^2,\alpha \mid y)\halfquad\propto\halfquad&(\sigma^2)^{-\frac{m}{2}-s-1}\exp\bigg\{-\frac{(\V{\bar{y}}-\V{\beta})\Tra W (\V{\bar{y}}-\V{\beta})+\operatorname{SSE}+2r}{2\sigma^2}\\&-\iota_{\mathcal{E}}(\V{\beta},\alpha)-(n-k+s_2)\log(1+\alpha)\bigg\},
\end{split}
\end{align}
where
\begin{eqnarray*}
\V{\bar{y}} &=& (\bar{y}_{1.},\bar{y}_{2.},\dots,\bar{y}_{n.})\Tra,\\
\M{W}&=&  \operatorname{diag}(w_1,w_2,\dots,w_n),\\
\operatorname{SSE}&=& \sum_{i=1}^n \sum_{j=1}^{w_i} (y_{ij}-\bar{y}_{i.})^2.
\end{eqnarray*}

There is no simple algorithm for projection onto $\mathcal{E}$ when $k\ge 1$. To take advantage of the epigraph projection algorithms described in \Sec{poe}, we consider the  reparameterization  $\V{\theta} =\M{T}_1\V{\beta}$ where
\begin{eqnarray}\label{eq:reparam1}
    \M{T}_1 = &\begin{bmatrix} \M{I}_{(k+1)\times n}\\ \M{D}^{(\V{x},k+1)}_n\end{bmatrix},
\end{eqnarray}
and $\M{I}_{(k+1)\times n}$ is the matrix obtained by taking the first $k + 1$ rows of a $n$-by-$n$ identity matrix.
In other words, we have $\V{\theta}_{[1:(k+1)]}=\V{\beta}_{[1:(k+1)]}$ and $\V{\theta}_{[(k+2):n]}=\M{D}^{(\V{x},k+1)}_n\V{\beta}$. To better visualize the reparameterization technique, we explicitly write out the reparameterization scheme for $x_i=i,\; i=1,2,\dots,n$ and $k=1$,
\begin{eqnarray*}
\begin{bmatrix}
\theta_1\\
\theta_2\\
\theta_3\\
\theta_4\\
\vdots\\
\theta_n
\end{bmatrix}& = &
\begin{bmatrix}
1 & 0 & 0 & 0 & \cdots & 0 & 0 & 0\\
0 & 1 & 0 & 0 & \cdots & 0 & 0 & 0\\
1 & -2 & 1 & 0 & \cdots & 0 & 0 & 0\\
0 & 1 & -2 & 1 & \cdots & 0 & 0 & 0\\
\vdots & & & & & & &\\
0 & 0 & 0 & 0 & \cdots & 1 & -2 & 1
\end{bmatrix}
\begin{bmatrix}
\beta_1\\
\beta_2\\
\beta_3\\
\beta_4\\
\vdots\\
\beta_n
\end{bmatrix}.
\end{eqnarray*}

Note that the transformation matrix $\M{T}_1$ is a lower-triangular banded matrix with $k+2$ non-zero diagonals. This means that given $\V{\theta}$, we can retrieve $\V{\beta}$ in $O(n(k+2))$ operations using a banded forward-solve step. The reparameterized posterior is
\begin{align}\label{eq:reparamposterior1}
\begin{split}
    \pi(\V{\theta},\sigma^2,\alpha \mid \V{y})\halfquad\propto\halfquad&(\sigma^2)^{-\frac{m}{2}-s-1}\exp\bigg\{-\frac{(\overline{\V{y}}-\M{T}_1^{-1}\V{\theta})\Tra \M{W}(\overline{\V{y}}-\M{T}_1^{-1}\V{\theta})+\operatorname{SSE}+2r}{2\sigma^2}\\&-\iota_{\mathcal{E}_1'}(\V{\theta},\alpha)-(n-k+s_2)\log(1+\alpha)\bigg\},
\end{split}
\end{align}
where 
\begin{eqnarray*}
\mathcal{E}_1' &=& \{(\V{\theta},\alpha)\in \Real^n\times \Real_{++} : \lVert \V{\theta}_{[(k+2):n]}\rVert_1\le\alpha\}.
\end{eqnarray*}
Replacing $\iota_{\mathcal{E}_1'}(\V{\theta},\alpha)$ with its Moreau-Yosida envelope, we arrive at a smooth surrogate posterior
\begin{align}\label{eq:reparamsurrogateposterior1}
\begin{split}
    \pi^\lambda(\V{\theta},\sigma^2,\alpha \mid \V{y})\halfquad\propto\halfquad&(\sigma^2)^{-\frac{m}{2}-s-1}\exp\bigg\{-\frac{(\overline{\V{y}}-\M{T}_1^{-1}\V{\theta})\Tra \M{W}(\overline{\V{y}}-\M{T}_1^{-1}\V{\theta})+\operatorname{SSE}+2r}{2\sigma^2}\\&-\frac{1}{2\lambda}d^2_{\mathcal{E}_1'}(\V{\theta},\alpha)-(n-k+s_2)\log(1+\alpha)\bigg\},
\end{split}
\end{align}

Projection onto $\mathcal{E}_1'$ can be accomplished by applying the $\ell_1$-norm epigraph projection process described in \Sec{poe} to $\V{\theta}_{[(k+2):n]}$. Working with this reparameterization raises some potential computational challenges, however. When evaluating the function value and calculating the gradient of \Eqn{reparamsurrogateposterior1}, we need to solve two linear systems, namely $\M{T}_1^{-1}\V{\theta}$ and $\M{T}_1^{-T}\M{W}(\overline{\V{y}}-\M{T}_1^{-1}\V{\theta})$. As $n$ and $k$ increases, the condition number of $\M{T}_1$ increases, leading to numerical instability in the HMC sampler. To alleviate this numerical issue, we can use the projection onto the epigraph of $\lVert \M{D}^{(1)}_n\V{\beta} \rVert_1$, described in \Sec{poe}. Borrowing the idea of \cite{ramdas2016fast}, we consider another reparameterization scheme $\V{\theta} =\M{T}_2\V{\beta}$ where 
\begin{eqnarray}\label{eq:reparam2}
    \M{T}_2 &=&\begin{bmatrix} \M{I}_{k\times n} \\\diag\left(\frac{k}{x_{k+1}-x_1},\dots,\frac{k}{x_{n}-x_{n-k}}\right)\M{D}^{(\V{x},k-1)}_n\end{bmatrix}.
\end{eqnarray}
The reparameterized density is now
\begin{align}\label{eq:reparamposterior2}
\begin{split}
    \pi(\V{\theta},\sigma^2,\alpha \mid \V{y})\halfquad\propto\halfquad&(\sigma^2)^{-\frac{m}{2}-s-1}\exp\bigg\{-\frac{(\overline{\V{y}}-\M{T}_2^{-1}\V{\theta})\Tra \M{W}(\overline{\V{y}}-\M{T}_2^{-1}\V{\theta})+\operatorname{SSE}+2r}{2\sigma^2}\\&-\iota_{\mathcal{E}'_2}(\V{\theta},\alpha)-(n-k+s_2)\log(1+\alpha)\bigg\},
\end{split}
\end{align}
where
\begin{eqnarray*}
\mathcal{E}'_2 &=& \{(\V{\theta},\alpha)\in \Real^n\times \Real_{++} : \lVert \M{D}^{(1)}_{n-k}\V{\theta}_{[(k+1):n]}\rVert_1\le\alpha\}.
\end{eqnarray*}
Similarly, projection onto $\mathcal{E}'$ can be achieved by applying the $\lVert \M{D}^{(1)}_n\V{\beta} \rVert_1$ epigraph projection process to $\V{\theta}_{[(k+1):n]}$. The advantage of using $\M{T}_2$ as the reparameterization scheme is that the ``order" of $\M{T}_2$ is one below that of $\M{T}_1$, so that solving the linear systems becomes more numerically stable. When $n$ and $k$ are relatively small, however, using $\M{T}_2$ requires solving \Eqn{fusedlasso}, which is more expensive than \Eqn{softthreshold}. \Tab{reparam} summarizes the approximate cutoffs of when to use $\M{T}_1$ and when to use $\M{T}_2$, based on our empirical studies. \Tab{reparam} does not include $k=0$ and $k=3$, since \cite{faulkner2018locally} demonstrated that the shrinkage property of the Laplace prior struggles to capture abrupt jumps of piecewise constant underlying trends, resulting in a posterior fit that is too wiggly. Our prior set up is analogous to the Laplace prior, so that our method runs into the same issue.  Meanwhile, when $k=3$, even $\M{T}_2$ is extremely ill-conditioned and the HMC sampler is hampered from exploring the parameter space meaningfully. Therefore, we focus on the case where $k=1$ (piecewise linear) and $k=2$ (piecewise quadratic).

Using $\M{T}_2$ as the reparameterization matrix mitigates but does not eliminate the ill-conditioning issue. As $n$ increases, it becomes more difficult for the HMC sampler to sufficiently explore the parameter space due to numerical instability. We will introduce a data preprocessing technique called thinning in \Sec{thinning} as an alternative strategy to make PBTF applicable for long sequences with large $n$.

\begin{table}[t]
\centering
\begin{tabular}{|c|c|c|c|}
\hline
      & $n\le 200$ & $200 <n\le 1000$ & $n>1000$       \\ \hline
$k=1$ & $\M{T}_1$  & $\M{T}_2$        & thinning needed \\ \hline
$k=2$ & $\M{T}_2$  & thinning needed   & thinning needed \\ \hline
\end{tabular}
\caption{Choice of reparameterization scheme for different $n$ and $k$.}
\label{tab:reparam}
\end{table}

Replacing $\iota_{\mathcal{E}'_2}(\V{\theta},\alpha)$ with its Moreau-Yosida envelope, the surrogate posterior is now
\begin{align}
\begin{split}\label{eq:reparamsurrogateposterior2}
    \pi^\lambda(\V{\theta},\sigma^2,\alpha \mid \V{y})\halfquad\propto\halfquad&(\sigma^2)^{-\frac{m}{2}-s-1}
    \exp\bigg\{-\frac{(\overline{\V{y}}-\M{T}_2^{-1}\V{\theta})\Tra \M{W}(\overline{\V{y}}-\M{T}_2^{-1}\V{\theta})+\operatorname{SSE}+2r}{2\sigma^2}\\&-\frac{1}{2\lambda}d^2_{\mathcal{E}'_2}(\V{\theta},\alpha)-(n-k+s_2)\log(1+\alpha)\bigg\}.
\end{split}
\end{align}
Notice that \Eqn{reparamsurrogateposterior1} and \Eqn{reparamsurrogateposterior2} are now differentiable functions of $(\V{\theta},\sigma^2,\alpha)$ on $\Real^n\times\Real_{++}\times\Real_{++}$.

For notational simplicity, in the rest of the manuscript we will use $\pi(\V{\theta},\sigma^2,\alpha \mid \V{y})$ to  refer to both \Eqn{reparamposterior1} and \Eqn{reparamposterior2},
$\pi^\lambda(\V{\theta},\sigma^2,\alpha \mid \V{y})$ to refer to and \Eqn{reparamsurrogateposterior1} and  \Eqn{reparamsurrogateposterior2}, and $(\M{T},\mathcal{E}')$ to refer to  $(\M{T}_1,\mathcal{E}_1')$ and  $(\M{T}_2,\mathcal{E}_2')$. We overload notation in this way since proofs and statements about these two surrogate densities are essentially the same. 

\subsection{Adding Shape-Restrictions}\label{sec:pbsrtf}
Proximal MCMC presents a simple alternative framework to traditional Bayesian hierarchical models that can easily construct priors that encode multiple structural constraints. Similar to nonparameteric isotonic trend filtering (\cite{kim2009ell_1, ramdas2016fast}), 
\begin{figure}[tbp]
  \centering
  \includegraphics[angle=0,width=\linewidth]{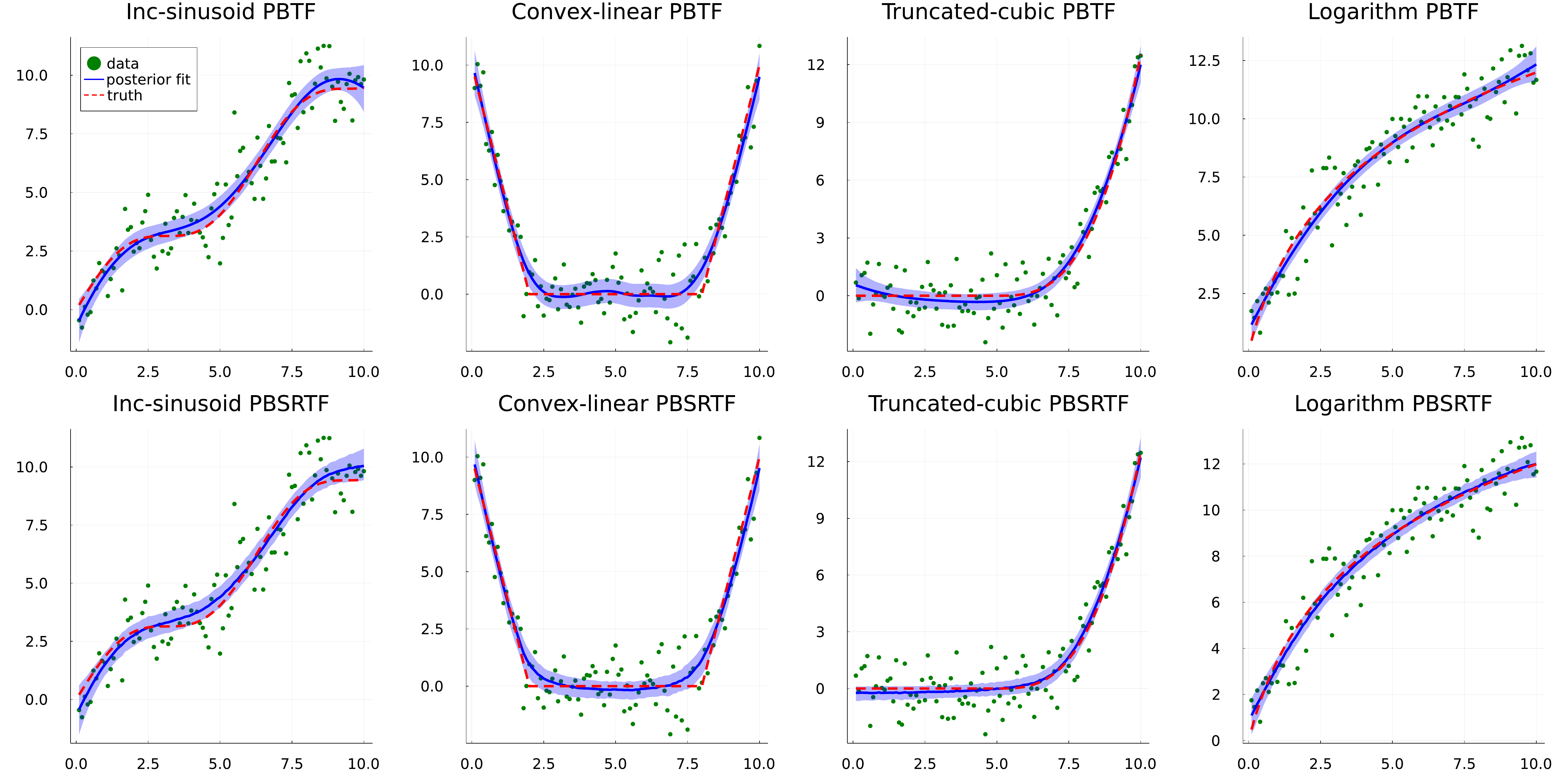}
  \caption{Example posterior fits for PBTF and PBSRTF with noise level $\sigma=1$.  The top row shows posterior fits of PBTF, and the bottom row shows posterior fits of PBSRTF. From left to right, the enforced shape restrictions are increasing, convex, increasing-convex and increasing-concave.}
  \label{fig:pbsrtfimage}
\end{figure}
adding shape restrictions into our framework is as straightforward as imposing linear inequalities.
For instance, if we believe that the underlying trend is monontone increasing, we can enforce monotonicity by refining the epigraph set $\mathcal{E}$ with a monotonicity constraint as follows
\begin{eqnarray*}
\mathcal{S}&=&\{(\V{\beta},\alpha)\in\Real\times\Real_{++} : \lVert \M{D}^{(\V{x},k+1)}_n\V{\beta}\rVert_1\le \alpha, \M{D}^{(1)}_n\V{\beta}\ge \V{0}\}.
\end{eqnarray*}

In addition to monotonicity, convexity can be encoded by the linear inequalities in \Eqn{convexity}. By replacing $\ge$ with $\le$, we get monotone decreasing and concave restrictions. Combining monotonicity and convexity is as simple as imposing two sets of linear inequalities. Therefore, our framework can model eight types of shape restrictions, namely increasing, decreasing, convex, concave, increasing-convex, increasing-concave, decreasing-convex and decreasing-concave. Lower or upper bounds on $\V{\beta}$ can also be enforced if warranted or desired. 

\Fig{pbsrtfimage} illustrates examples of posterior fits using both versions of our fully Bayesian proximal MCMC method for trend filtering with and without shape-restrictions. 
For proof of concept, projection onto $\mathcal{S}$ can be achieved by any quadratic programming solver. We report the results using the Gurobi solver and leave for future work developing customized algorithms for potentially greater scalability.

As alluded to earlier, for PBSRTF we consider a joint prior on $(\V{\beta},\alpha)$ that employs a Gamma prior on $\alpha$
\begin{eqnarray}\label{eq:jointgammasr}
    \pi(\V{\beta},\alpha)&\propto&\exp\{-\iota_{\mathcal{S}}(\V{\beta},\alpha)-\mu\alpha\}.
\end{eqnarray}
The joint prior in \Eqn{jointgammasr} is almost identical to the one in \Eqn{jointgamma}; we simply replaced $\mathcal{E}$ with $\mathcal{S}$, where shape restrictions are also present. There are several reasons to revisit a Gamma prior on $\alpha$. First, we can no longer interpret $\mathcal{S}$ as an $\ell_1$-norm ball so that it is unclear what the normalizing constant should be; contrast this to the non shape-restricted case where the normalizing constant is $\alpha^{-(n-k-1)}$. In fact, using $\alpha^{-(n-k-1)}$ as the normalizing constant for PBSRTF results in too much shrinkage. Second, there are numerical challenges that make the sampler using the beta-prime prior typically slower overall. We discuss these challenges in the supplementary materials. Finally, issues of the posterior being sensitive to the choice of $\mu$, as we highlighted in \Sec{pbtf}, are no longer  prohibitively acute as in the non shape-restricted case. In the case of PBSRTF, shape restrictions impose a helpful dose of regularization on $\V{\beta}$, therefore blunting the influence of our choice of $\mu$. 

Using an inverse Gamma $\operatorname{IG}(s,r)$ as the prior for $\sigma^2$ and \Eqn{jointgammasr} as the prior for $(\V{\beta},\alpha)$, our full posterior density for PBSRTF is 
\begin{align}\label{eq:fullposteriorpbsrtf}
\begin{split}
    \pi(\V{\beta},\sigma^2,\alpha \mid \V{y})\halfquad\propto\halfquad&(\sigma^2)^{-\frac{m}{2}-s-1}\exp\left\{-\frac{(\V{\bar{y}}-\V{\beta})\Tra \M{W} (\V{\bar{y}}-\V{\beta})+\operatorname{SSE}+2r}{2\sigma^2}-\iota_{\mathcal{S}}(\V{\beta},\alpha)-\mu\alpha\right\}.
\end{split}
\end{align}
Replacing $\iota_{\mathcal{S}}(\V{\beta},\alpha)$ with its Moreau-Yosida envelope,  results in the surrogate posterior
\begin{align}\label{eq:fullposteriorsurrogatepbsrtf}
\begin{split}
    \pi^\lambda(\V{\beta},\sigma^2,\alpha\mid \V{y})\halfquad\propto\halfquad&(\sigma^2)^{-\frac{m}{2}-s-1}\exp\left\{-\frac{(\V{\bar{y}}-\V{\beta})\Tra \M{W} (\V{\bar{y}}-\V{\beta})+\operatorname{SSE}+2r}{2\sigma^2}-\frac{1}{2\lambda}d_{\mathcal{S}}^2(\V{\beta},\alpha)-\mu\alpha\right\}.
\end{split}
\end{align}
Again, \Eqn{fullposteriorsurrogatepbsrtf} is a differentiable function of $(\V{\beta},\sigma^2,\alpha)$ on $\Real^n\times\Real_{++}\times\Real_{++}$. Neither \Eqn{reparamsurrogateposterior2} nor  \Eqn{fullposteriorsurrogatepbsrtf} is log-concave, however, so that Langevin algorithms are no longer suitable for MCMC sampling. Therefore, we turn to Hamiltonian Monte Carlo as our sampling engine.

\subsection{Properties of the Surrogate Posteriors}

We conclude this section, with two theorems that justify the practice of replacing the nonsmooth part of the posterior by its Moreau-Yosida envelope. The proofs are provided in the supplementary materials.
\begin{theorem}\label{thm:propriety}
The surrogate posterior densities \Eqn{reparamsurrogateposterior1},\Eqn{reparamsurrogateposterior2} and \Eqn{fullposteriorsurrogatepbsrtf} are proper, i.e.,
\begin{eqnarray*}
\int_{\Real^n}\int_{\Real_{++}}\int_{\Real_{++}}\pi^\lambda(\V{\theta},\sigma^2,\alpha\mid\V{y})d\V{\theta}d\sigma^2d\alpha &<& +\infty,
\end{eqnarray*}
and
\begin{eqnarray*}
\int_{\Real^n}\int_{\Real_{++}}\int_{\Real_{++}}\pi^\lambda(\V{\beta},\sigma^2,\alpha\mid\V{y})d\V{\beta}d\sigma^2d\alpha &<&  +\infty.
\end{eqnarray*}
\end{theorem}
\begin{theorem}\label{thm:convergence}
The surrogate posterior densities \Eqn{reparamsurrogateposterior1},\Eqn{reparamsurrogateposterior2} and \Eqn{fullposteriorsurrogatepbsrtf} converges to the original nonsmooth densities \Eqn{reparamposterior1},\Eqn{reparamposterior2} and \Eqn{fullposteriorpbsrtf} in total-variation norm as $\lambda\downarrow 0$, i.e.,
\begin{eqnarray*}
\underset{\lambda\downarrow 0}\lim\int_{\Real^n}\int_{\Real_{++}}\int_{\Real_{++}}\left\lvert\pi^\lambda(\V{\theta},\sigma^2,\alpha\mid\V{y})-\pi(\V{\theta},\sigma^2,\alpha\mid\V{y})\right\rvert d\V{\theta}d\sigma^2d\alpha &= &0,
\end{eqnarray*}
and
\begin{eqnarray*}
\underset{\lambda\downarrow 0}\lim\int_{\Real^n}\int_{\Real_{++}}\int_{\Real_{++}}\left \lvert \pi^\lambda(\V{\beta},\sigma^2,\alpha\mid\V{y})-\pi(\V{\beta},\sigma^2,\alpha\mid\V{y})\right\rvert d\V{\beta}d\sigma^2d\alpha &=&0.
\end{eqnarray*}
\end{theorem}
\Thm{convergence} assures us that the surrogate density can approximate the  original posterior density arbitrarily well by choosing a small enough $\lambda$. This is consistent with our experiments where we observe that the posterior fit is visually smooth once $\lambda$ is sufficiently small. Note that $\lambda$ should not be chosen to be too small, however, as doing so will lead to numerical instability since gradient evaluations involve division by $\lambda$. We discuss how to properly choose $\lambda$ for the two different applications in in the supplementary materials. In practise, we recommend using the default parameters in our software.
%% ----------------------------------------------------------------------
%% Computation
%% ----------------------------------------------------------------------
\section{Posterior Computation}
\label{sec:computation}
\subsection{HMC Sampling}
We apply Hamiltonian Monte Carlo (HMC) to sample from the smoothed surrogate full posterior densities \Eqn{reparamsurrogateposterior1},\Eqn{reparamsurrogateposterior2} and \Eqn{fullposteriorsurrogatepbsrtf}. Software for the proposed method is available at \url{https://github.com/qhengncsu/ProxBTF.jl}. We implement our method with \textsl{DynamicHMC.jl} package in the \textsl{Julia} computing environment. According to its documentation, the package implements a variant of the ``No-U-Turn Sampler" (NUTS) of \cite{hoffman2014no}, as described in \cite{betancourt2017conceptual}. We direct readers to \cite{betancourt2017conceptual} for an accessible exposition on the algorithmic details of the sampling scheme. Since the NUTS algorithm operates on an unrestricted domain, we reparameterize $\sigma^2$ as $e^{\log\sigma^2}$ and $\alpha$ as $e^{\log\alpha}$ to model the two positive parameters. 

For PBTF, evaluating the function-gradient pair at any given location requires $O(n)$ operations. While using Gurobi as a black box solver obscures the computational complexity of PBSRTF, we observe empirically that the computation time of PBSRTF also scales linearly with grid length $n$. This is likely due to the fact that Gurobi can effectively exploit the sparse matrices in our problem set up.

\subsection{Thinning}\label{sec:thinning}
As discussed in \Sec{pbtf}, PBTF may encounter numerical difficulties that accompany solving ill-conditioned linear systems. While the difference epigraph projection technique alleviates the ill-conditioning issue, it can not eliminate it; as $n$ increases, eventually the condition number of $\M{T}_2$ will eventually become problematic.

Another technique we propose to mitigate the ill-conditioning issue is thinning, which is similar to the thinning practice in R package \textsl{glmgen} \cite{ramdas2016fast}. 
We first split the range of $\V{x}$ into intervals of equal length. Grid locations within the same interval are merged into a single new grid location, which is a weighted average of the original grid locations with weights being the numbers of observations. The data points $(x_i,y_{ij})$ are then horizontally shifted to the merged grid locations. After HMC sampling, if we are interested in the posterior median and confidence limits at the original grid locations, we can recover them through interpolation. We provide an illustration of thinning in the supplementary materials.

%% ----------------------------------------------------------------------
%% Experiments
%% ----------------------------------------------------------------------

\section{Numerical Experiments}\label{sec:pbtfvsspmrfs}
We compare the empirical performance of PBTF with Shrinkage Prior Markov Random Fields (SPMRFs) by \cite{faulkner2018locally} and Dynamic Shrinkage Processes (DSP) by \cite{kowal2019dynamic}. We note that DSP can be considered as an extention of SPMRFs and the software of DSP\footnote{https://github.com/drkowal/dsp} in fact contains an implementation of the hierarchical models described in \cite{faulkner2018locally}. Moreover, DSP uses customized Gibbs samplers which in practice are more efficient than the HMC sampler used by SPMRFs, thus we primarily use the software of DSP in our experiments. In \Tab{pbtftable}, BTF-BL (Bayesian Lasso prior or Laplace prior) and BTF-HS (horseshoe prior) correspond to the models presented in \cite{faulkner2018locally} while BTF-DHS (dynamic horseshoe prior) corresponds to the model presented in \cite{kowal2019dynamic}. To investigate the relative strengths of different approaches, we selected four underlying trends, namely piecewise linear, smooth trend, sinusoid, and piecewise quadratic/cubic. We assess the precision of each method with mean absolute deviation (MAD), frequentist coverage probability (CP), and mean credible interval width (MCIW). We also include the total CPU time (TCPU), effective sample size of the slowest component (min.\@ ESS) and multivariate effective sample size (MESS) \citep{vats2019multivariate} as measures of sampling efficiency. The detailed definitions of the summary statistics are given in the supplementary materials.
\begin{table}[htbp]
\centering
{\footnotesize
\begin{tabular}{@{}cccccccc@{}}
\toprule
True Trend                             & Method       & MAD (s.d.)  & MCIW & CP   & TCPU(s) &min. ESS & MESS \\ \midrule
\multirow{4}{*}{Piece. Linear}         & BTF-BL       & 0.87 (0.18) & 4.3  & 0.95 & 12      &2271  & 4018 \\
                                       & BTF-HS       & 0.73 (0.19) & 3.7  & 0.95 & 7       &1368  & 3275 \\
                                       & BTF-DHS      & 0.70 (0.18) & 3.7  & 0.95 & 17      &880   & 3120 \\
                                       & PBTF ($k=1$) & 0.82 (0.17) & 3.9  & 0.94 & 70      &1902  & 2037 \\ \midrule
\multirow{4}{*}{Smooth Trend}          & BTF-BL       & 0.98 (0.16) & 5.1  & 0.96 & 12      &1674  & 2440 \\
                                       & BTF-HS       & 1.00 (0.15) & 5.1  & 0.95 & 7       &973   &2491 \\
                                       & BTF-DHS      & 1.02 (0.15) & 5.1  & 0.95 & 17      &150   &1893 \\
                                       & PBTF ($k=2$) & 0.87 (0.16) & 4.3  & 0.95 & 896     &857   &2684 \\ \midrule
\multirow{4}{*}{Sinusoid}              & BTF-BL       & 0.80 (0.14) & 4.6  & 0.97 & 12      &2080  &4120 \\
                                       & BTF-HS       & 0.83 (0.14) & 4.7  & 0.97 & 7       &1203  &2340 \\
                                       & BTF-DHS      & 0.86 (0.14) & 4.8  & 0.97 & 17      &260   &1884 \\
                                       & PBTF ($k=2$) & 0.70 (0.14) & 3.9  & 0.97 & 927     &1207  &3686 \\ \midrule
\multirow{4}{*}{Piece. Quad./ Cubic}   & BTF-BL       & 0.77 (0.12) & 4.3  & 0.97 & 12      &845   &4223 \\
                                       & BTF-HS       & 0.78 (0.15) & 4.1  & 0.96 & 7       &378   &2585 \\
                                       & BTF-DHS      & 0.82 (0.15) & 4.2  & 0.95 & 17      &180   &2190 \\
                                       & PBTF ($k=2$) & 0.70 (0.13) & 3.8  & 0.96 & 931     &1439  &3326 \\ \bottomrule
\end{tabular}}
\caption{Summary statistics for DSP and PBTF, averaged over 50 generated sequences at noise level $\sigma=3$.}
\label{tab:pbtftable}
\end{table}

Following \cite{faulkner2018locally} and \cite{kowal2019dynamic}, we used evenly spaced grid locations of  $\{1,2,\dots,100\}$ and designed the underlying trends to have an approximate standard deviation of 9. We added two levels of Gaussian noise ($\sigma=3.0$ and $\sigma=4.5$) to the underlying trends, generating 50 noisy sequences for each combination of trend and noise level. For DSP, we used the default parameters, ran an initial burn-in of 1000 iterations followed by 2500 posterior draws. For PBTF, we set $s_2$ to be $\sqrt{n}=10$, ran the default warm-up stage in \textsl{DynamicHMC.jl} and made another 2500 posterior draws. \Tab{pbtftable} shows the summary statistics for different methods averaged over 50 generated sequences with $\sigma=3.0$. The results for noise level $\sigma=4.5$ can be found in the supplementary materials, which exhibits a similar pattern.

The last three trends, namely smooth trend, sinuoid and piecewise quadratic/cubic are better approximated by piecewise quadratic functions. However, in \Tab{pbtftable} we only report the results of DSP using $k=1$. This is partly because the software of DSP does not contain an option to fit models with $k=2$. That being said, the software of SPMRFs\footnote{https://github.com/jrfaulkner/spmrf} does offer an option to fit models with $k=2$. Nevertheless, for the last three trends, when going from $k=1$ to $k=2$, SPMRFs overall suffers a decrease in MAD and CP in contrary to one's expectation. These additional simulation results can be found in the supplementary materials. SPMRFs' worse performance with $k=2$, despite the underlying trends being better approximated by piecewise quadratic functions, may be attributed to the fact that it is inherently harder to sample from higher-order trend filtering models. In our framework, third-order PBTF alleviates part of that difficulty through leveraging the fused lasso subroutine, providing the best MAD and the narrowest confidence bands for the last three trends while maintaining ideal coverage probability. 

BTF-HS achieves higher precision than BTF-BL and PBTF for piecewise linear trend, demonstrating stronger adaptivity to abrupt turns. This is attributed to the superior shrinkage properties of global-local priors like the horseshoe prior. Unfortunately, nonparametric analogues of the horseshoe prior are nonconvex, for example, smoothly clipped absolute deviation (SCAD) penalty \citep{fan2001variable} and minimax concave penalty (MCP) \citep{zhang2010nearly}. Projection onto the epigraph of a nonconvex function is generally nontrivial. Therefore, it is not immediately obvious how to replicate the horseshoe prior's shrinkage property in our framework and presents an interesting avenue for future work. BTF-DHS achieved even better precision than BTF-HS for piecewise linear trend through modelling dependence between the local scale parameters. However, we also see that it will behave slightly worse than BTF-HS when modelling smooth underlying trends.

We note that DSP only applies to data on an evenly spaced grid. The framework of SPMRFs is extended to unevenly spaced grids for $k=0$ and $k=1$ in \cite{faulkner2018locally} using methods based on integrated Wiener processes. However, \cite{faulkner2018locally} did not further pursue the same for $k=2$ due to its complexity. PBTF, on the other hand, naturally handles unevenly spaced grids for $k=1,2$ due to using the adjusted difference matrix $\M{D}^{(\V{x},k+1)}_n$ in its prior. In this section, we employed an evenly spaced grid $\{1,2,\dots,n\}$ in pursuit of simplicity and conformity. The real data analysis in \Sec{real} and the thinning example in the supplementary materials are both examples of third-order PBTF being applied to unevenly spaced grids. 

\section{Real Data Example}\label{sec:real}
\begin{figure}[htbp]
  \centering
  \includegraphics[width=\linewidth]{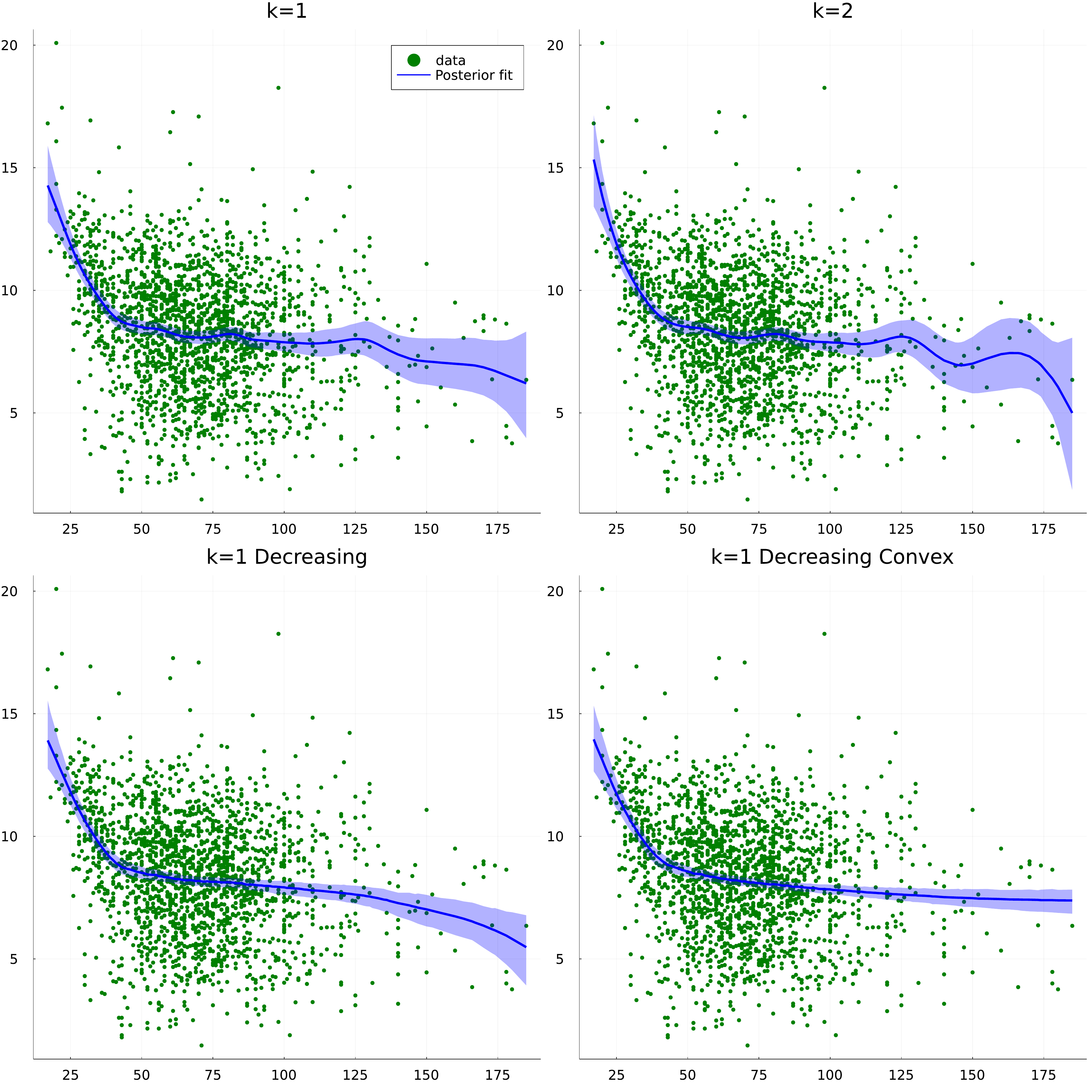}
  \caption{Posterior fits on Munich dataset. Plots show data points (green dots), posterior median (blue solid lines), and 95\% Bayesian credible intervals (light blue bands).}
  \label{fig:munich}
\end{figure}
We apply PBTF and PBSRTF to the Munich dataset as a real data example. We focus on two variables in the dataset, with the response being rent per square meter in Munich, Germany, and the covariate being floor space in square meters. The dataset was first analysed by \cite{rue2005gaussian} using Gaussian Markov Random Fields. \cite{faulkner2018locally} analyzed this data as an illustration of SPMRFs being applied on an unevenly spaced grid. The dataset has $2035$ observations in total and the covariate floor space has 134 distinct values. Other than second-order and third-order PBTF models, we also present second-order PBSRTF model fits with ``decreasing" and ``decreasing-convex" as shape restrictions. In the former case, we model the assumption that rent per square meter decreases as floor space increases. In the latter case, we model an additional diminishing returns effect.

We used $s_2=2\times\sqrt{134}$ for PBTF and set $\mu=4.0$ for PBSRTF to promote a bit more regularity. \Fig{munich} shows the posterior fits of the four different models. All four models captured an overall decreasing trend. It is notable that the confidence bands are narrower over intermediate values of floor space, which is expected as there are more data points over this range of floor spaces. Third-order PBTF produced a more variable posterior median and a wider confidence band than second-order PBTF, suggesting that third-order PBTF models exhibit more adaptivity but may be prone to overfitting.  We notice that posterior fits with shape restrictions have much narrower confidence bands compared with their unconstrained counterparts. This is because the shape restrictions introduce additional regularization that further reduces variance.

%% ----------------------------------------------------------------------
%% Discussion
%% ----------------------------------------------------------------------
\section{Discussion}
In this work, we introduced a new proximal MCMC methodology, which incorporates the variance parameter $\sigma^2$ and the regularization parameter $\alpha$ into posterior inference. The key to extending the conventional proximal MCMC paradigm to a fully Bayesian one is to use epigraph priors to induce sparsity and regularity. By substituting the nonsmooth components of the posterior with its Moreau-Yosida envelope, we can work with a differentiable surrogate density, on which HMC is be applied for efficient MCMC sampling.

As a proof of concept, we explored the application of the proposed methodology in Bayesian trend filtering. Compared with existing Bayesian trend filtering methods, our approach achieves higher precision for underlying trends that are better approximated by piecewise quadratic functions. To demonstrate the flexibility of our framework, we also explored incorporating shape restrictions like monotonicity and convexity. 

Although we focused on Bayesian trend filtering in this work, 
the strategy of combining an epigraph prior with proximal MCMC readily applies to other types of nonsmooth estimation problems. For example,  modern optimization extensively utilizes nuclear norms to induce low-rank structure, therefore a Bayesian version of low-rank matrix completion based on projection onto the epigraph of nuclear norm is an interesting future venue. It is also of great appeal to venture beyond convex penalties and constraints for greater modelling power in structured regression problems.

\begin{center}
    {\large \textbf{Supplementary Materials}}
\end{center}
\begin{description}
\item[Title:] Supplement to ``Bayesian Trend Filtering via Proximal Markov Chain Monte Carlo". (.pdf file)
\item[Software:] Julia-package “ProxBTF.jl” containing
code to perform the methods described in the article and scripts (R and Julia) to reproduce the numerical experiments. (.zipped file)
\end{description}
%\begin{center}
    %{\large \textbf{Acknowledgement}}
%\end{center}
%We thank the editor, the associate editor and the anonymous referee for inspiring comments and helpful suggestions, which greatly improved the presentation of this paper. We thank James Faulkner for helpful discussions. This research was partially funded by grants from the National Institute of General Medical Sciences (R35GM141798: HZ, R01GM135928: EC), the National Human Genome Research Institute (R01HG006139: HZ), and the National Science Foundation (DMS-2054253 and IIS-2205441: HZ, DMS-2201136: EC). Much of the motivation for this work originated from talks on proximal MCMC methods by Marcelo Pereyra at a pair of workshops on operator splitting methods in data science which were funded by the National Science Foundation under Grant (DMS-1127914, the Statistical and Applied Mathematical Sciences Institute) and the Simons Foundation. We thank Marcelo for stimulating discussion that sparked interest in pursuing this work. We thank Patrick Combettes and Christian M\"{u}ller for organizing these workshops as well as Ilse Ipsen for her oversight in planning the first workshop as the SAMSI Directorate Liaison.
%\begin{center}
    %{\large \textbf{Disclosure Statement}}
%\end{center}
%The authors report there are no competing interests to declare.

\bibliographystyle{asa}
\bibliography{refs}

\end{document}